\documentclass[12pt]{article}
\usepackage[english]{babel}
\usepackage{amsmath,amsthm,amssymb,amsfonts,latexsym}
\usepackage{color}

\usepackage{graphicx}
\usepackage{subfigure}
\catcode `\@=11 \@addtoreset{equation}{section}

\catcode `\@=12

\begin{document}

\title{Political dynamics affected by turncoats}  

\author{R.~D.~Salvo, M.~Gorgone and F.~Oliveri\\
\ \\
{\footnotesize Department of Mathematical and Computer Sciences,}\\
{\footnotesize Physical Sciences and Earth Sciences, University of Messina}\\
{\footnotesize Viale F. Stagno d'Alcontres 31, 98166 Messina, Italy}\\
{\footnotesize rdisalvo@unime.it; mgorgone@unime.it; foliveri@unime.it}
}

\date{Published in \textit{Int. J. Theor. Phys.} \textbf{56}, 3604--3614 (2017).}

\maketitle

\begin{abstract}
An operatorial theoretical model based on raising and lowering fermionic operators for the description of the dynamics of a political system consisting of macro--groups affected by turncoat--like behaviors is presented. 
The analysis of the party system dynamics is carried on by combining the action of a suitable quadratic Hamiltonian operator with specific rules (depending on the variations of the mean values of the observables) able to adjust periodically the conservative model to the political environment.
\end{abstract}


\section{Introduction}
\label{sect1}
The dynamics of a political system where different parties are involved is influenced and/or determined by many factors of difficult characterization. Nevertheless, in recent years, 
the mathematical modelization of some aspects, such as the creation of coalitions or cooperations between different political parties, political decision making, voting rules and vote maximizers, attracted the attention of many researchers.
A broader insight into some questions and methods of the mathematical approaches which are used in political science, together with comparative static and dynamic path predictions, can be found in \cite{Johnson_Book}. 
Also, either classical epidemiological approaches or typical game theory tools have been used to build mathematical models for the analysis of spread of political parties, as well as of coalition formation and  change of voters' opinion over time due to 
self--reflection, communication, random external events, and noise, respectively (see 
\cite{Lichtenegger,Misra,Sened}, and references therein). 
Moreover, the methods of quantum information theory, and the formalism of the theory of open quantum systems have been used to describe the dynamics of the voters' mental states and the approaching of a stable state of a decision equilibrium \cite{Khrennikova_Haven,Khrennikova}.

In this paper, we aim to model the dynamics of a political system whose parties are characterized by the tendency of part of their members (referred to as \emph{turncoats}) to change even repeatedly allegiance. 
The model is built by using an operatorial approach typical of quantum mechanics \cite{Merzbacher,Roman} in the Heisenberg picture, suitably extended \cite{BDSGO_Hrho2016}.  
As shown in  \cite{BagarelloBook}, the description of the dynamics of macroscopic complex 
systems may be profitably carried out by using  \emph{raising} and \emph{lowering} operators and the number representation;  in fact, this method has been used in several recent papers to analyze the dynamical aspects in rather different areas
\cite{BagarelloBook,BagarelloOliveriMigration2013,BagarelloHaven2014a,BagarelloHaven2014b,BagarelloOliveriEco2014,BagarelloGarganoOliveriEscape2015,BagarelloCherubiniOliveriDesert2016,DiSalvoOliveriRM2016,DiSalvoOliveriAAPP2016}. 
Moreover, operatorial techniques have been used to set up dynamical models describing alliances in politics \cite{BagarelloAlliance2015,BagarelloHavenAlliance2016} with  reference to the
Italian political system. 

Fermionic annihilation and creation operators, together with their associated number operators \cite{Merzbacher,Roman}, are used to define local densities of observables of a macroscopic political system affected by ``turncoat--like'' behaviors. Figuratively speaking, a turncoat is a person who changes allegiance from one loyalty or ideal to another one in an unscrupulous way to get maximum benefits and personal gain. The reason of the shift is frequently self--interest, and the faithless attitude in politics may come upon in a manifest way through an open and declared change of side, or in a veiled and subtle way in terms of incoherence in the voting secrecy (just think of those politicians from fragmented subfactions who declare a certain ideology, but
secretly give electoral support to a different cause). 
The effects of such a behavior in a political system are somehow similar to those 
occurring in the classic Prisoner's Dilemma, where players choose independently whether to contribute toward a mutually desirable outcome. Contributing is optimal for the group and suboptimal for the individual. Thus not contributing is a dominant strategy, which is to say, standard game theory predicts defection. Even in the case of an $n$--person Prisoner's Dilemma \cite{game1}, no matter what strategy the other players are playing, it is better for an individually rational player to defect than to cooperate. But each person's payoff in the equilibrium profile is lower than each person's payoff in the profile in which all cooperate, also in the profiles with mixed cooperation and defection. Hence, individual rationality does not lead to collective rationality, and the presence of turncoats can do a lot of damage to others.
In the study of voting systems \cite{game2}, the possibility of cheating is considered as one alternative strategy that may or may not be followed according to the range of possible interacting outcomes. 
In real world, cheating is never an emotionally or morally neutral choice (so that almost every language is full of pejorative terms for defectors); nevertheless, with respect to the contemporary political landscape in place in Italy, a look at the behavior of politicians from various parties during the first thirty months of the XVII Italian Legislature (started in 2013) reveals a high level of 
disloyal attitude and openness towards accepting chameleons
from other political groups in both houses of the Italian Parliament. Such kind of ugly phenomena, 
which manifest in more than 300 changes of party during the period of 30 months under observation, clearly emerge from the official data available in the institutional web pages of the Chamber of Deputies 
({http://www.camera.it}) and the Senate of the Republic ({http://www.senato.it}). In \cite{DSO_turncoat}, a special case of the general model described in this paper has been presented as a case study, and it has been shown that  the numerical results of the operatorial model there considered give a satisfactory fit with the
real data. Moreover, in \cite{DSGO_voters}, within the same theoretical framework adopted in this paper, the influence of the opportunistic attitudes of politicians on voters' opinion has been investigated. 

The model considered in this paper is suitable to describe a political system made by $N$ political parties each thought of as composed by $N$ subfactions, and an additional coalition (called \emph{mixed group}) 
made by politicians who refused to be members of any other party. Therefore, we deal with $N^2+1$  fermionic modes, so that the Hilbert space in which the system lives (see Section~\ref{sect2}) will result finite dimensional. The dynamics of such a model is ruled by a 
time--independent self--adjoint quadratic Hamiltonian operator. 
Nevertheless, in order to obtain more realistic time evolutions, we assume the dynamics 
to be driven, besides the Hamiltonian, by the periodic action of a \emph{rule} able to
modify the values of some of the parameters on the basis of the state variations of the system.  

The plan of the paper is the following one. 
In Section~\ref{sect2}, we briefly sketch the operatorial framework, and introduce the model. We also describe how to derive the dynamics by using the standard Heisenberg representation together with the action of the \emph{rule}: this is what has been called 
$(H,\rho)$--induced dynamics \cite{DiSalvoOliveriAAPP2016,BDSGO_QGOL2016,BDSGO_Hrho2016}. 
In Section~\ref{sect3}, as an application, we consider the case $N=4$, and  present some numerical simulations.

\begin{figure}
\includegraphics[width=0.8\textwidth]{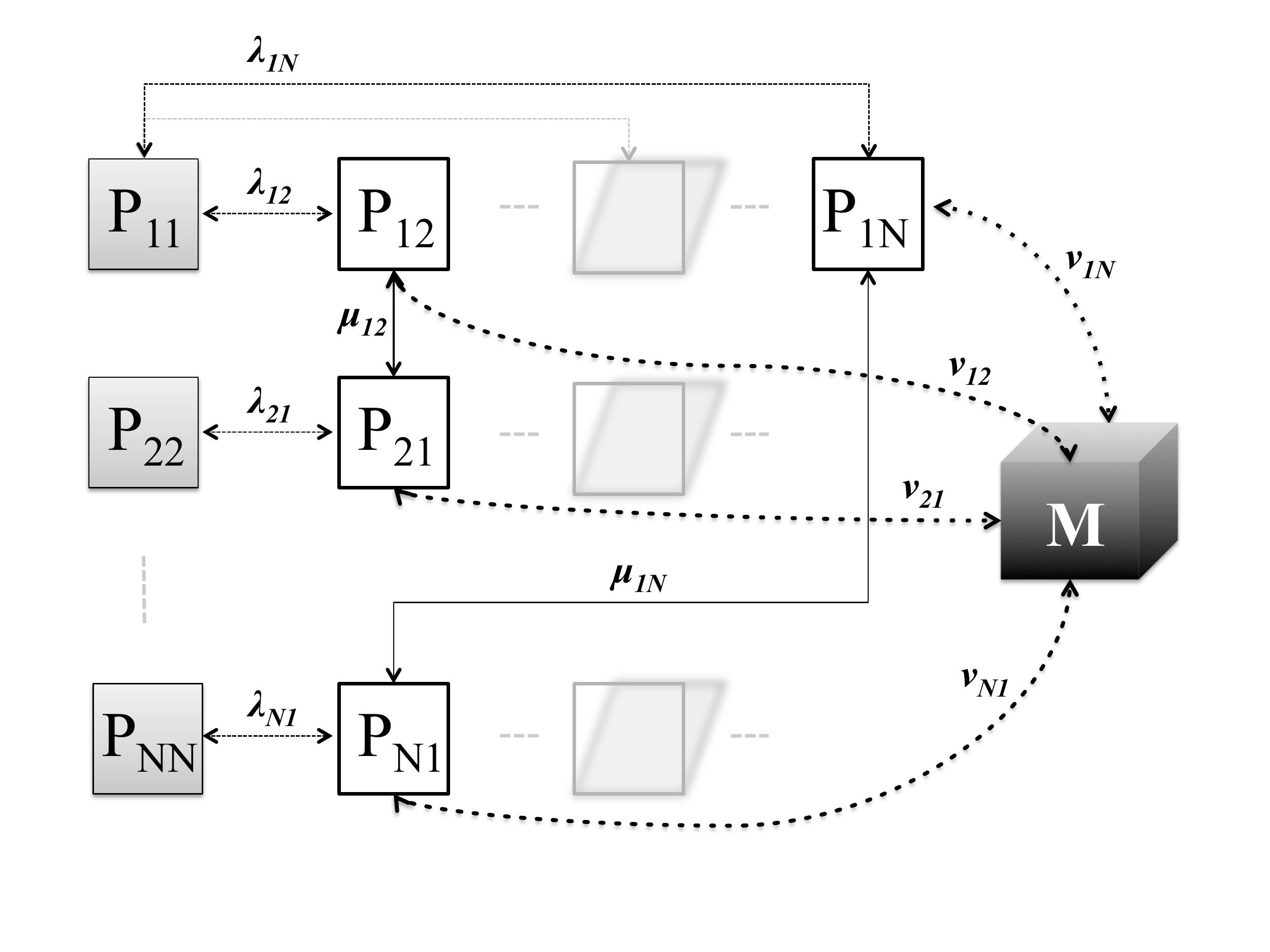}
\caption{\label{fig:schema}Schematic view of the model.}
\end{figure}

\section{Changes of party and allegiance: the model}
\label{sect2}

Let us consider a political system made up of $N$ different parties $\mathcal{P}_i$, each one thought of as composed by a loyal subgroup $\mathcal{P}_{ii}$, and $N-1$ disloyal fringes $\mathcal{P}_{ij}$,  
$(i,j=1,\ldots,N,\; j\neq i)$, and a further coalition $\mathcal{M}$, the
\emph{mixed group}, collecting politicians with heterogeneous ideologies. 
The politicians in the subfaction $\mathcal{P}_{ij}$ are members of the $i$--th party but at
the same time may be tempted to switch to the subfaction $\mathcal{P}_{ji}$ of the $j$--th party, or to the group $\mathcal{M}$, and vice versa, as schematized in Figure~\ref{fig:schema}. 

We associate to the $N^2+1$ compartments  annihilation ($P_{ij}$ and $M$) and 
creation ($P^\dagger_{ij}$ and $M^\dagger$) fermionic operators satisfying the \emph{Canonical Anti--commutation Rules} 
(CAR) 
\begin{equation}
\begin{aligned}
&\{P_{ij},P^\dagger_{kl}\}=\delta_{i,k}\delta_{j,l}\mathbb{I}, \qquad \{P_{ij},M^\dagger\}=0,\\
&\{P_{ij}, P_{kl}\}=\{P_{ij},M\}=\{P^\dagger_{ij}, P^\dagger_{kl}\}=\{P_{ij}^\dagger,M^\dagger\}=0,
\end{aligned}
\end{equation}
where
\[
\{X,Y\}:=XY+YX.
\]
 
The states of the system (involving a finite number of fermionic states) are vectors in the $2^{N^2+1}$--dimensional Hilbert
space $\mathcal{H}$ constructed as the linear span of the vectors 
\[
\Phi_{p_{11},p_{12},\ldots,p_{NN},m}:=(P_{11}^\dagger)^{p_{11}}(P_{12}^\dagger)^{p_{12}}\cdots(P_{NN}^\dagger)^{p_{NN}}(M^\dagger)^m\Phi_{\bf 0},
\]
which therefore turn out to form an orthonormal basis for $\mathcal{H}$, where $p_{ij},m\in \{0, 1\}$ for all $i,j=1, \ldots, N$, 
and the vector $\Phi_\mathbf{0}$ is the vacuum of the theory, \emph{i.e.}, a vector annihilated by all the operators 
$P_{ij}$ and $M$.

The vectors $\Phi_{p_{11},p_{12},\ldots,p_{NN},m}$ give an orthonormal set of eigenstates of the number operators 
$\widehat p_{ij}=P^\dagger_{ij}P_{ij}$, $\widehat m=M^\dagger M$, say,
\begin{equation}
\begin{aligned}
&\widehat p_{ij} \Phi_{p_{11},p_{12},\ldots,p_{NN},m}=p_{ij} \Phi_{p_{11},p_{12},\ldots,p_{NN},m},\\ 
&\widehat m \Phi_{p_{11},p_{12},\ldots,p_{NN},m}=m \Phi_{p_{11},p_{12},\ldots,p_{NN},m}.
\end{aligned}
\end{equation}

We assume at first that the dynamics of the system is ruled by the
self--adjoint time--independent Hamiltonian operator
\begin{equation}\label{eq:Ham_turn}
\begin{aligned}
H= &\sum_{i,j=1}^{N}\,\omega_{ij}\, P_{ij}^\dagger\,P_{ij} + \omega_{N+1}\, M^\dagger\,M+
\sum_{\substack{i,j=1\\ j\neq i}}^N \lambda_{ij}\left(P_{ii}P_{ij}^\dagger+P_{ij}P_{ii}^\dagger\right)\\
&+\sum_{1\le i<j\le N} \mu_{ij}\left(P_{ij}P_{ji}^\dagger+P_{ji}P_{ij}^\dagger\right)+
\sum_{\substack{i,j=1\\ j\neq i}}^N \nu_{ij}\left(P_{ij}M^\dagger+MP_{ij}^\dagger\right)
\end{aligned}
\end{equation}
embedding the main effects deriving from the interactions among the $N^2+1$ actors of the system, and has the property of 
determining the time evolution of any actor $X$ of the model, 
in the so--called Heisenberg representation, in which  the operators acting on the Hilbert space evolve in time  
according to the equation  $ \dot X(t)=\mathrm{i}[H,X(t)]=\mathrm{i}(HX(t)-X(t)H)$.

The real constants $\omega_{ij}$ are 
related to the tendency of each degree of freedom to stay constant in time; 
the larger their values,
the smaller the amplitudes of the oscillations of the related densities 
\cite{BagarelloBook}; in some sense, they are a measure of the \emph{inertia} of the various compartments. Moreover, the real parameters $\lambda_{ij}$  are used to describe the internal flows that occur within each faction, namely the minor ideological positions representing in some sense early warning of disloyalty, whereas the parameters $\mu_{ij}$ are related to the external flows stemming from the real changes of side; finally, the parameters $\nu_{ij}$ are related to the flows from and to the mixed group. Notice that, requiring $\lambda_{ij}=\mu_{ij}=\nu_{ij}=0$ $(i,j=1,\ldots,N)$,  meaning that the components of the system do not  interact, implies that 
$[H,\widehat p_{ij}] = [H,\widehat m]=0$: even if the operators $P_{ij}$, $M$, and $P^\dagger_{ij}$, $M^\dagger$ have a non--trivial  time dependence, the densities of the compartments of the system stay constant in time.

In accordance with the Heisenberg scheme, the linear equations of motion for the operators $P_{ij}$ and $M$ read
\begin{equation}
\label{eqs_Heis}
\left\{
\begin{aligned}
\dot P_{ii}(t)&=\mathrm{i} \left(-\omega_{ii} P_{ii}(t)+\sum_{\substack{k=1\\k\neq i}}^N \lambda_{ik} P_{ik}(t)\right),\\
\dot P_{ij}(t)&=\mathrm{i} \left(-\omega_{ij} P_{ij}(t)+\lambda_{ij} P_{ii}(t)+\mu_{k\ell} P_{ji}(t)+\nu_{ij} M(t)\right),\\
\dot M(t)&=\mathrm{i} \left(-\omega_{N+1} M(t)+\sum_{\substack{i,k=1\\k\neq i}}^N \nu_{ik} P_{ik}(t)\right),
\end{aligned}
\right.
\end{equation}
where $i,j=1,\ldots,N$, $j\neq i$, and $\displaystyle{\mu_{k\ell}=\begin{cases}\mu_{ij} &\mbox{if } i<j\\ \mu_{ji} &\mbox{otherwise} \end{cases}}$.

Since the total densities of political parties commute with $H$, say
\begin{equation}
\left[H,\left(\sum_{i,j=1}^N\hat{p}_{ij}\right)+\hat{m}\right]=0,
\end{equation}
that is the quantity $\left(\sum_{i,j=1}^N p_{ij}(t)\right)+m(t)$ is a constant, the system is suitable to describe the density conservation of the total number of political members.

The time evolution of any observable $X$ of the system (in particular, we are interested to
the number operators) is given by
\begin{equation}
X(t)=\exp(iHt)X\exp(-iHt),
\end{equation}
whereupon we compute its mean value
\begin{equation}\label{add3}
x(t)=\langle\Phi_{p_{11},p_{12},\ldots,p_{NN},m},\,
X(t)\Phi_{p_{11},p_{12},\ldots,p_{NN},m}\rangle
\end{equation}
on a given initial eigenstate.
The evolution of the density $\mathfrak{p}_i$ of each party $\mathcal{P}_i$ ($i=1,\ldots,N$) is easily obtained, say 
\begin{equation}
\mathfrak{p}_i(t)= \sum_{j=1}^{N} p_{ij}(t).
\end{equation}
The dynamics we may deduce with the Hamiltonian \eqref{eq:Ham_turn}
is at most quasiperiodic. More interesting behaviors can be obtained by means of the
so called $(H,\rho)$--induced dynamics \cite{DiSalvoOliveriAAPP2016,DSO_turncoat,BDSGO_QGOL2016,BDSGO_Hrho2016}. By considering a self--adjoint quadratic Hamiltonian operator $H^{(1)}\equiv H$, and the evolution of a certain observable $X$,
\begin{equation}
X(t)=\exp(iH^{(1)}t)X\exp(-iH^{(1)}t),
\end{equation}
we compute its mean value by using \eqref{add3} in a time interval of length $\tau>0$ on a vector $\Phi_{p_{11},p_{12},\ldots,p_{NN},m}$. Then, we modify some of the parameters (by a \emph{rule}) involved in $H^{(1)}$ on the basis of the values of the various $x(\tau)$. In this way, we get a new Hamiltonian operator $H^{(2)}$, having the same functional form as $H^{(1)}$, but with different values of the parameters. We now continue to follow the evolution of the system as ruled by $H^{(2)}$ for a second  time interval, again of length $\tau>0$. And so on.
As a result, the global evolution will be governed by a sequence of similar Hamiltonian operators, and the involved parameters are stepwise constant.

In general terms, let us consider a time interval $[0,T]$, and split it in $n=T/\tau$ subintervals of length $\tau$. Assume  $n$ to be integer. In the $k$--th subinterval $[(k-1)\tau,k\tau]$ consider a Hamiltonian $H^{(k)}$ ruling the dynamics. The global dynamics arises from the sequence of Hamiltonians
\begin{equation}
H^{(1)} \stackrel{\tau}{\longrightarrow} H^{(2)} \stackrel{\tau}{\longrightarrow} H^{(3)} \stackrel{\tau}{\longrightarrow} \ldots \stackrel{\tau}{\longrightarrow} H^{(n)},
\end{equation}
the complete evolution being obtained by gluing the local evolutions.

The fact that the parameters are repeatedly modified according to the actual state of the system produces
a self--adjustment of the model since the strength of the interactions  may change during the evolution; as a result, the model results able to mimic specific effects which it is not easy at all to include \emph{a priori} in a time--dependent $H$, and asymptotic values of the mean values can be obtained \cite{BDSGO_Hrho2016}.

Stated differently, by adopting a rule, we are implicitly considering the possibility of having a time--dependent Hamiltonian (in the sense that we get qualitatively similar evolutions, \cite{BDSGO_Hrho2016}). But the time--dependence is, in our case, of a very special form: $H$ is actually piecewise constant in time, \emph{i.e.}, in each interval $[(k-1)\tau,k\tau[$ the Hamiltonian does not depend on time, but in $k\tau$ some changes may occur, according to how the system is evolving. In addition, since in each subinterval the Hamiltonian is time--independent, the computations for deriving the dynamics are simpler than those required in the presence of a time--dependent Hamiltonian.  

\section{Numerical simulations}
\label{sect3}

\begin{figure}
\begin{center}
\subfigure[]{\label{a1}\includegraphics[width=0.44\textwidth]{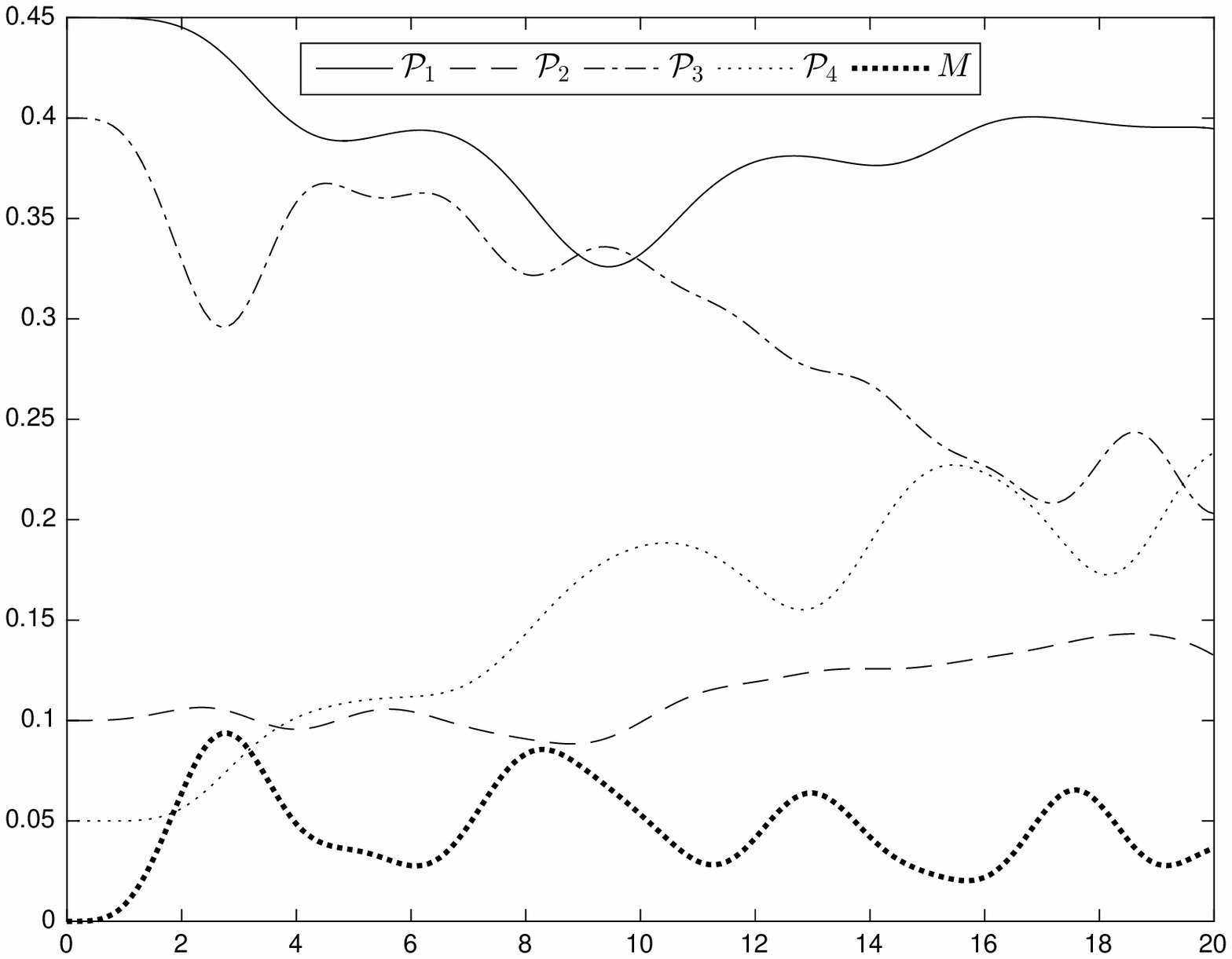}}\\
\subfigure[$\mathcal{P}_1$]{\label{b1}\includegraphics[width=0.44\textwidth]{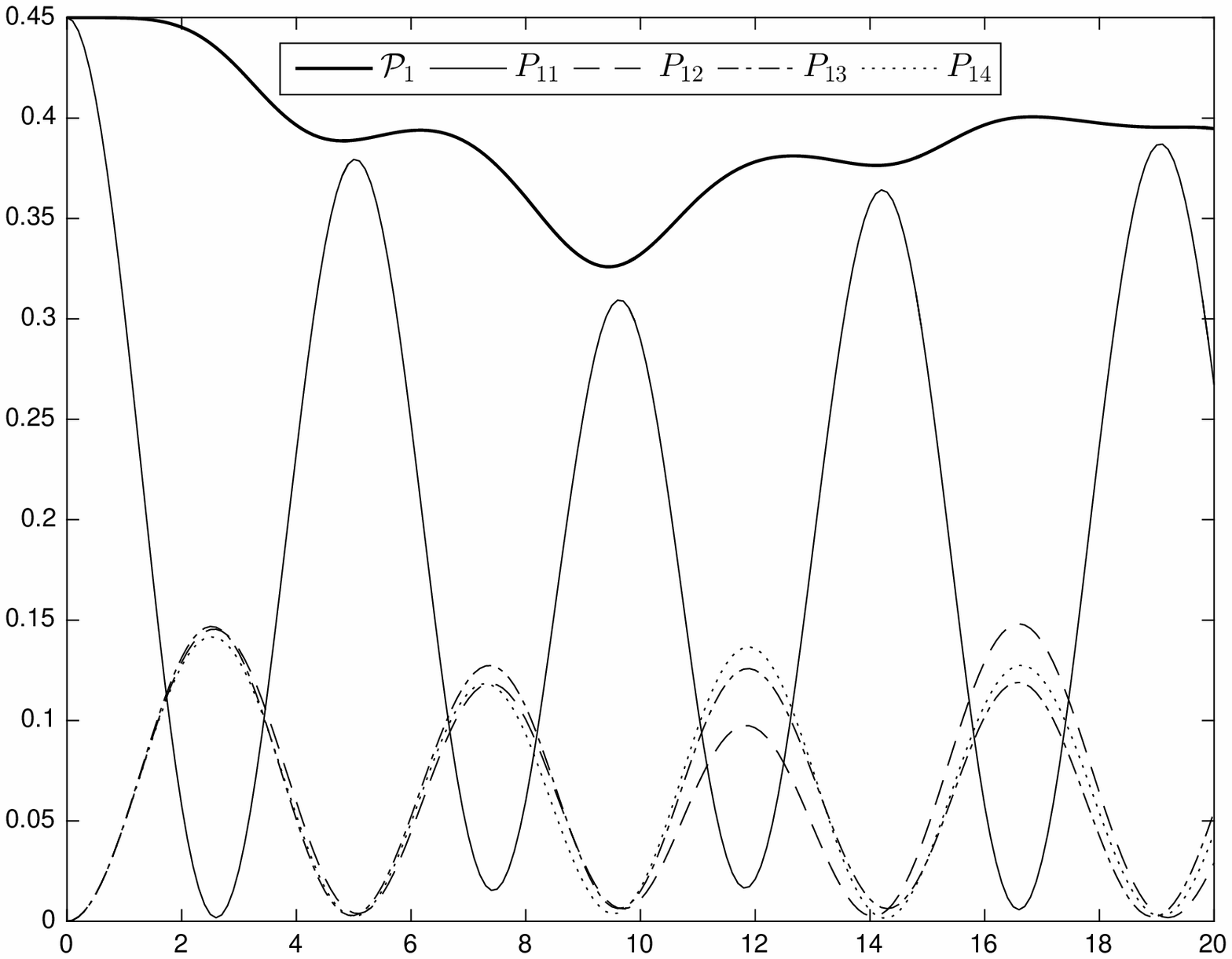}}\quad 
\subfigure[$\mathcal{P}_2$]{\label{c1}\includegraphics[width=0.44\textwidth]{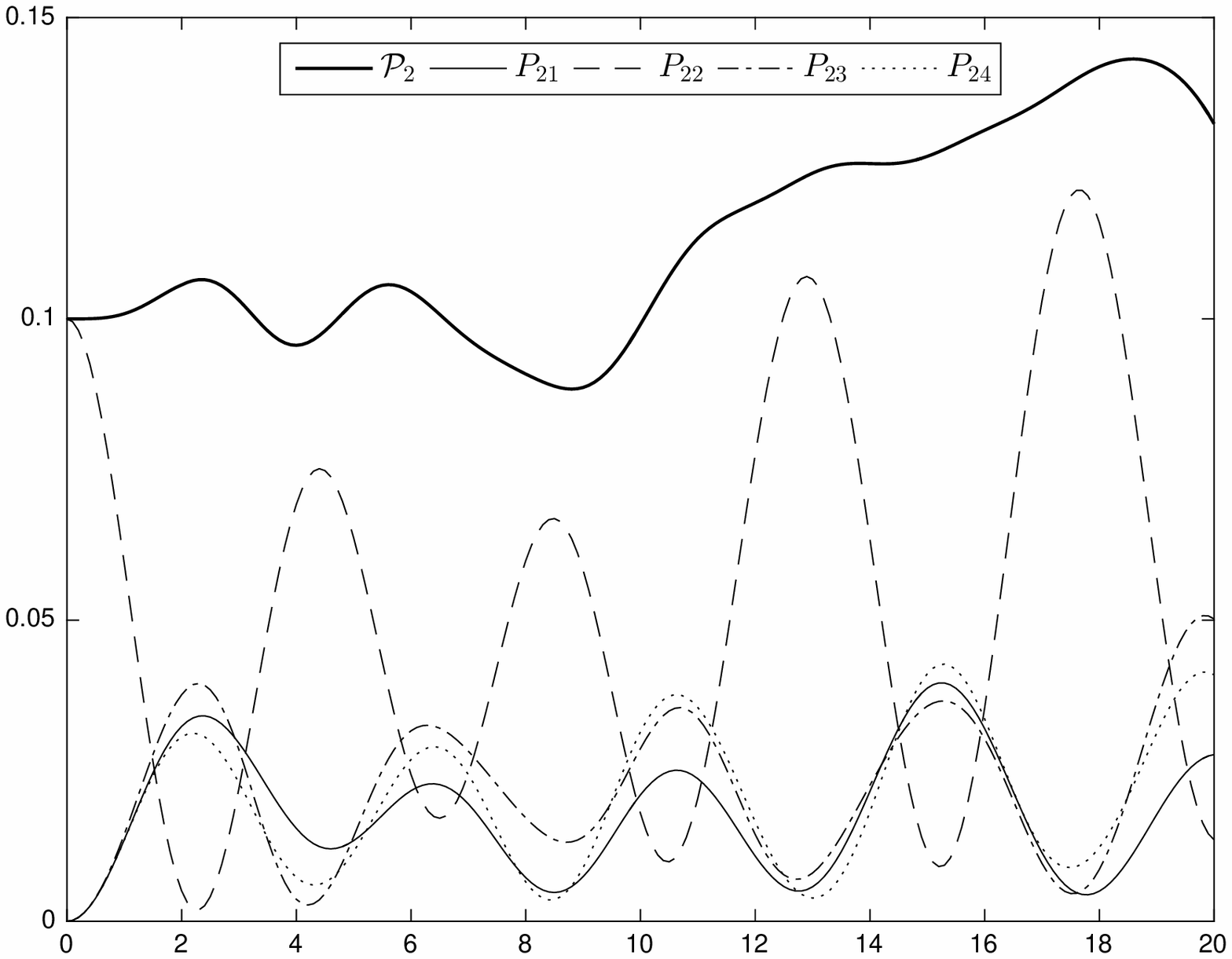}}\\
\subfigure[$\mathcal{P}_3$]{\label{d1}\includegraphics[width=0.44\textwidth]{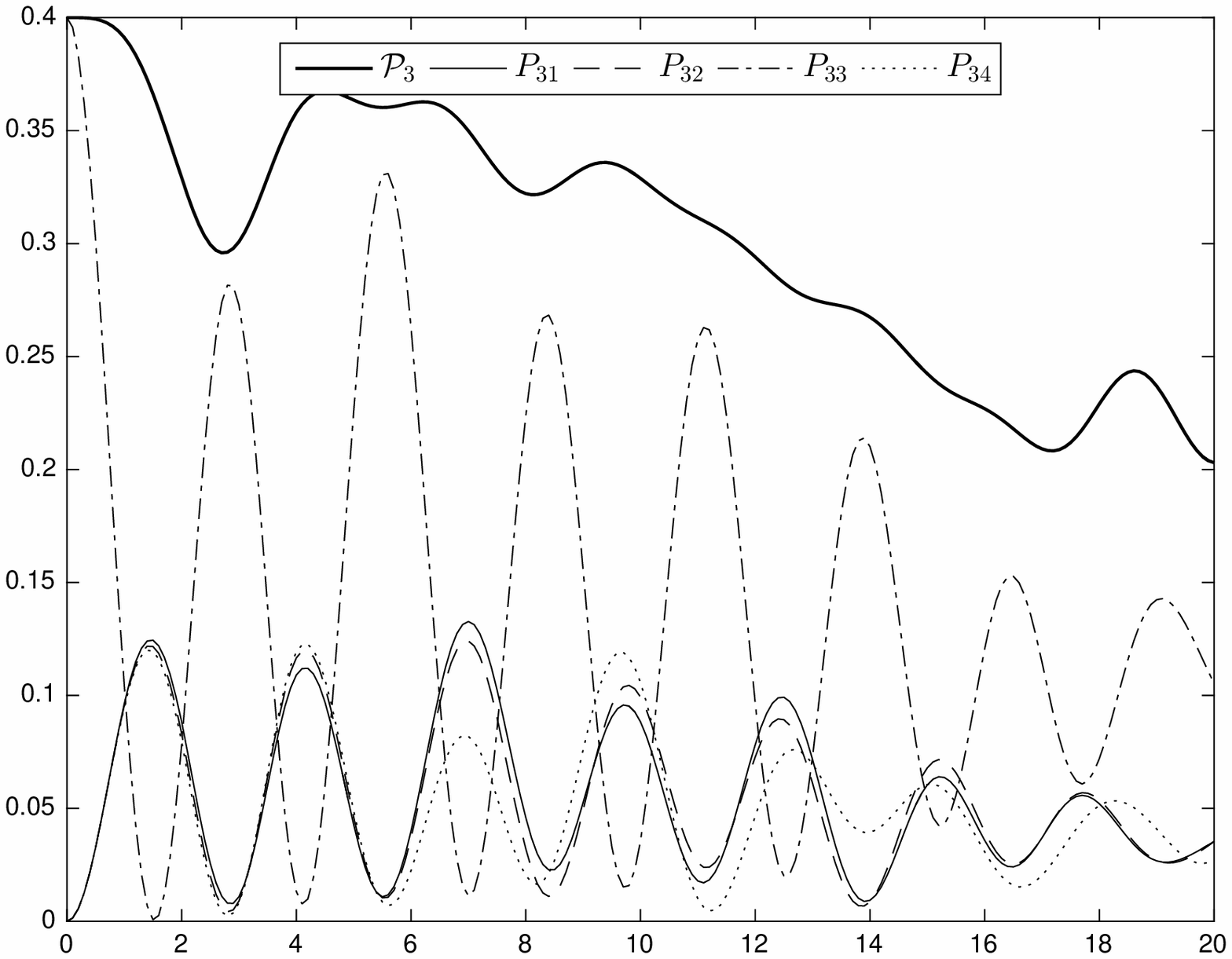}}\quad 
\subfigure[$\mathcal{P}_4$]{\label{e1}\includegraphics[width=0.44\textwidth]{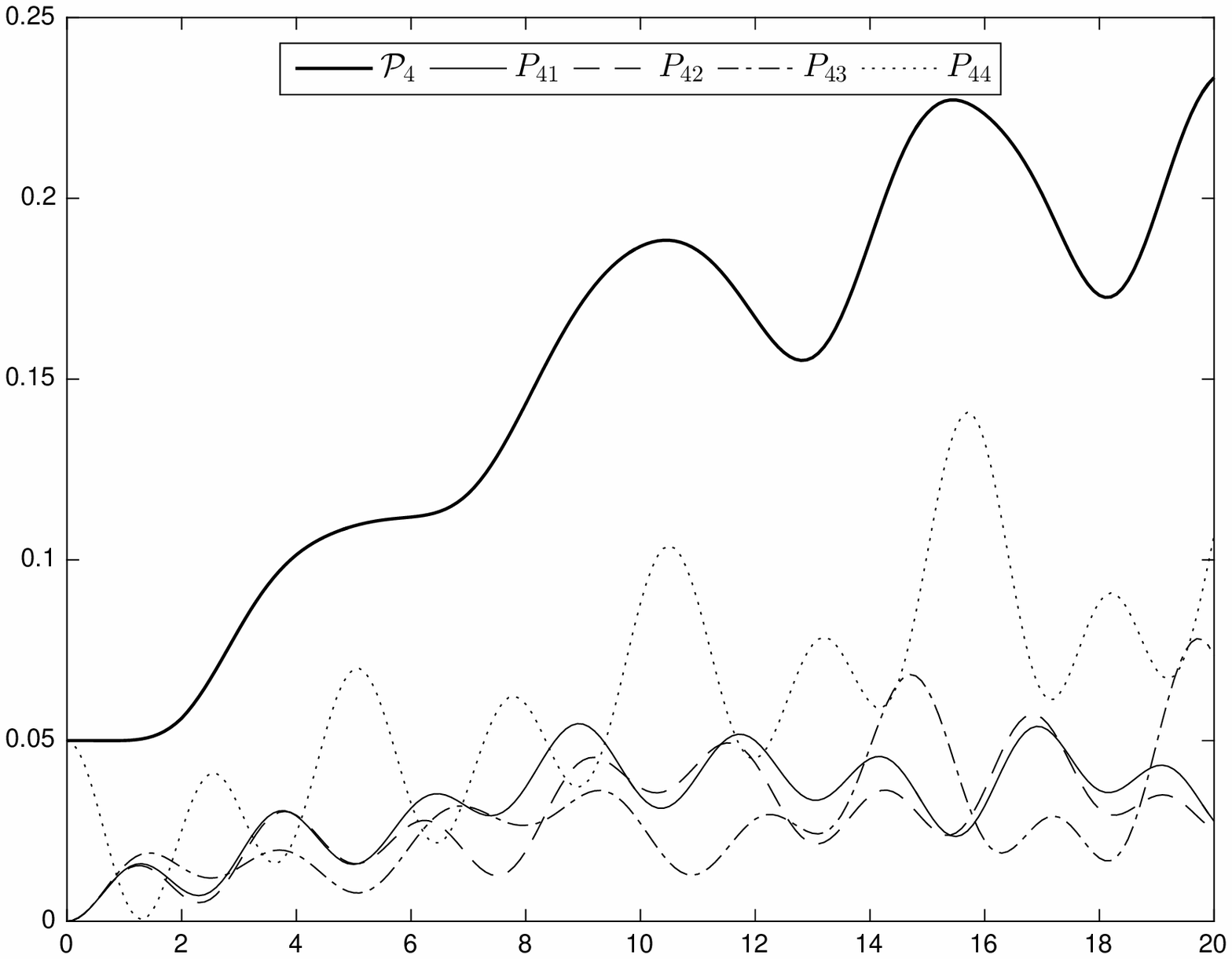}}
\caption{\footnotesize{\label{fig2} Dynamics of political parties affected by turncoat--like behaviors: the time evolution of the densities of the parties in \ref{a1} reproduces the reinforcement of the moderate factions. The frames \ref{b1}, \ref{c1}, \ref{d1}, \ref{e1} show the fluctuations of the densities of the various subgroups of each party. The $x$--axis is scaled to the number of quarters in a five--year term of office; the rule $\rho$ is applied with $\tau=1$.}}
\end{center}
\end{figure}

\begin{figure}
\begin{center}
\subfigure[]{\label{a2}\includegraphics[width=0.44\textwidth]{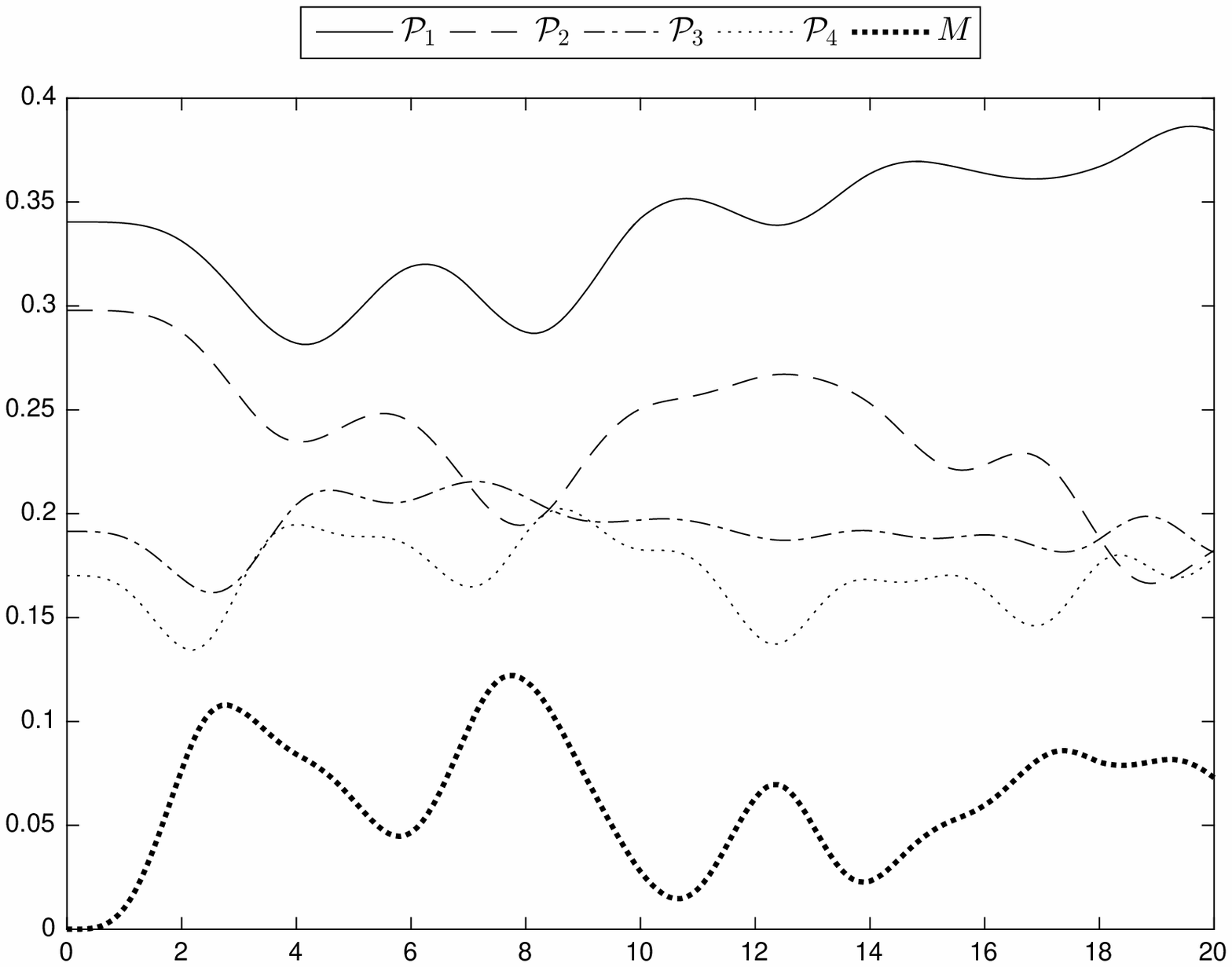}}\\
\subfigure[$\mathcal{P}_1$]{\label{b2}\includegraphics[width=0.44\textwidth]{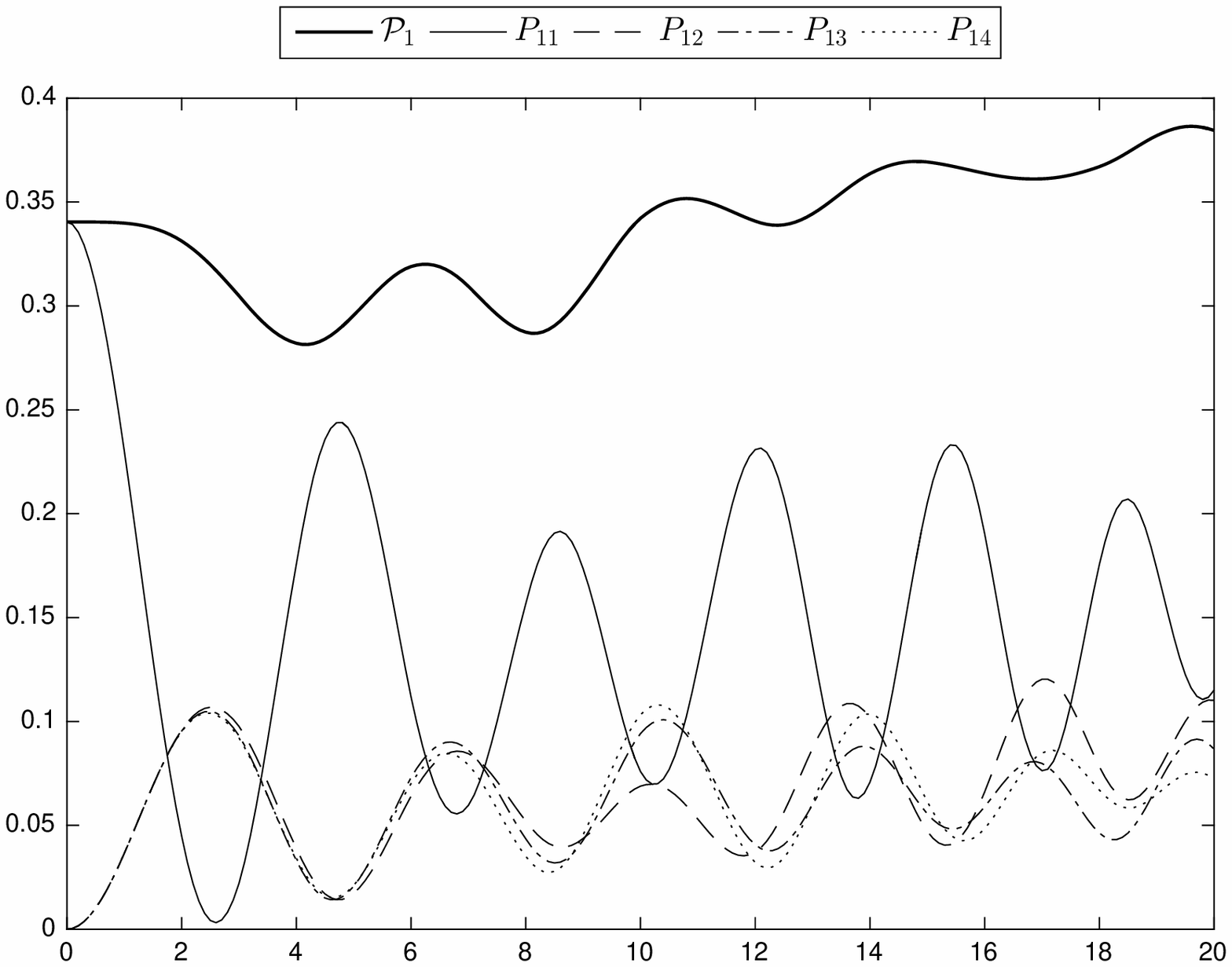}}\quad 
\subfigure[$\mathcal{P}_2$]{\label{c2}\includegraphics[width=0.44\textwidth]{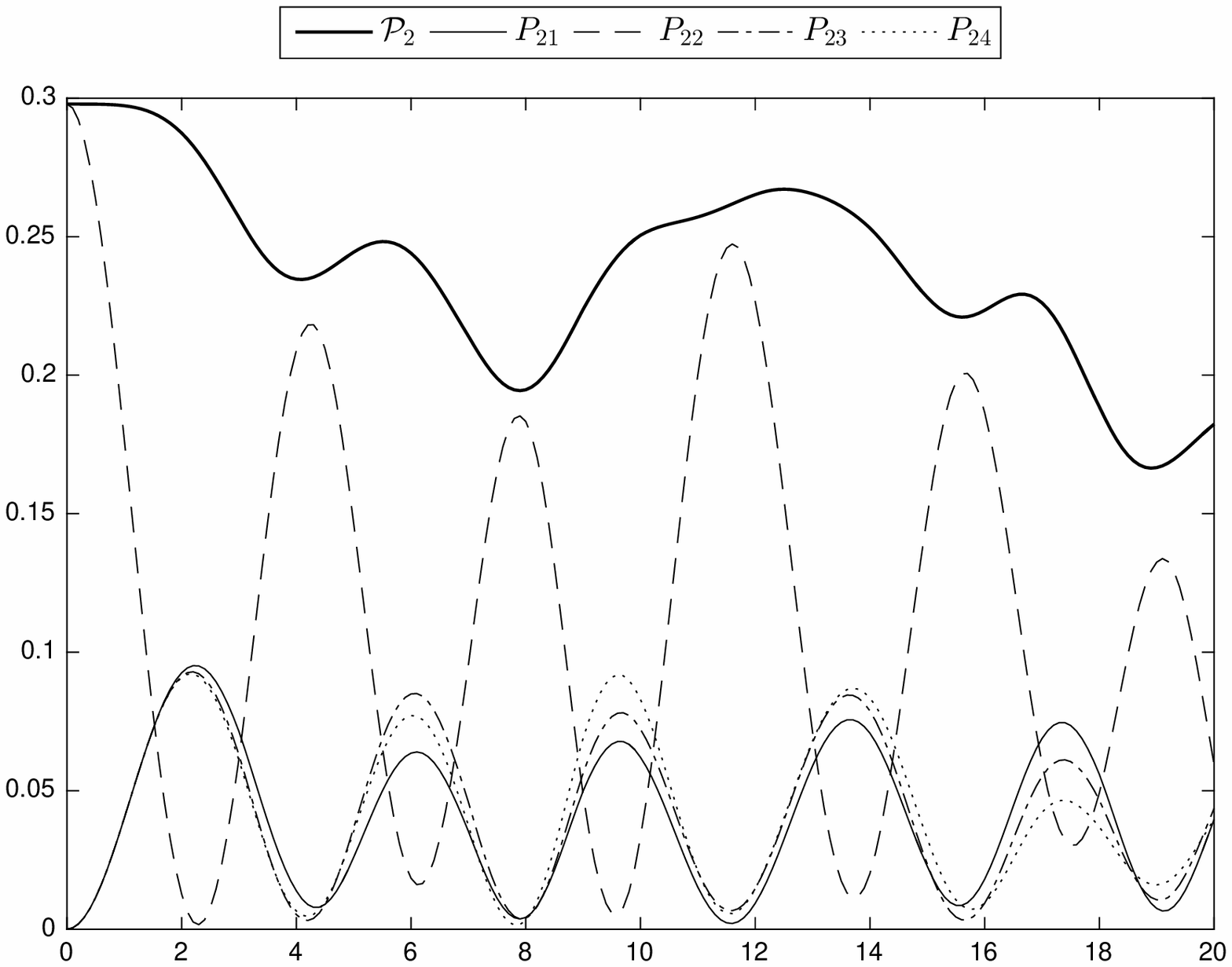}}\\
\subfigure[$\mathcal{P}_3$]{\label{d2}\includegraphics[width=0.44\textwidth]{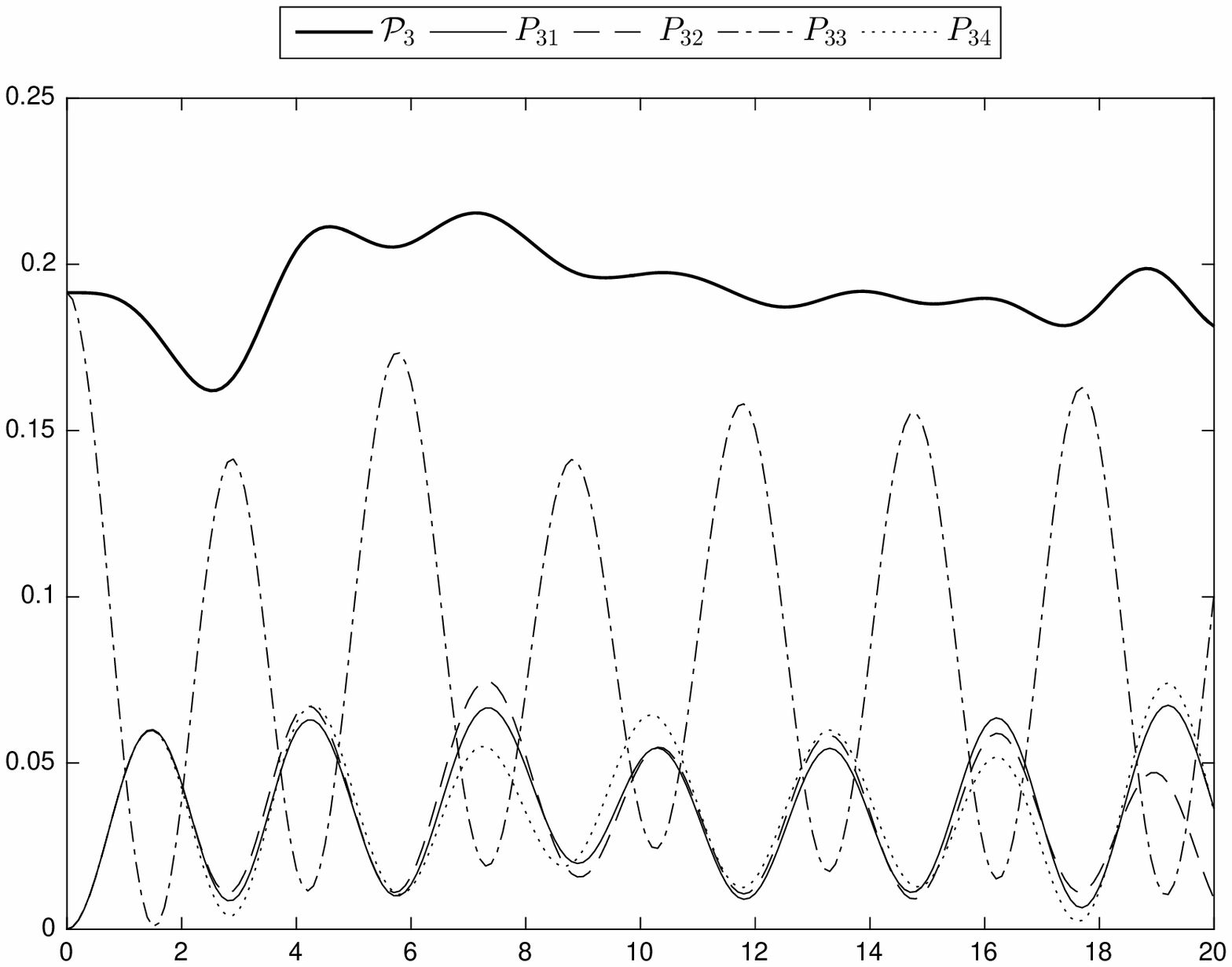}}\quad 
\subfigure[$\mathcal{P}_4$]{\label{e2}\includegraphics[width=0.44\textwidth]{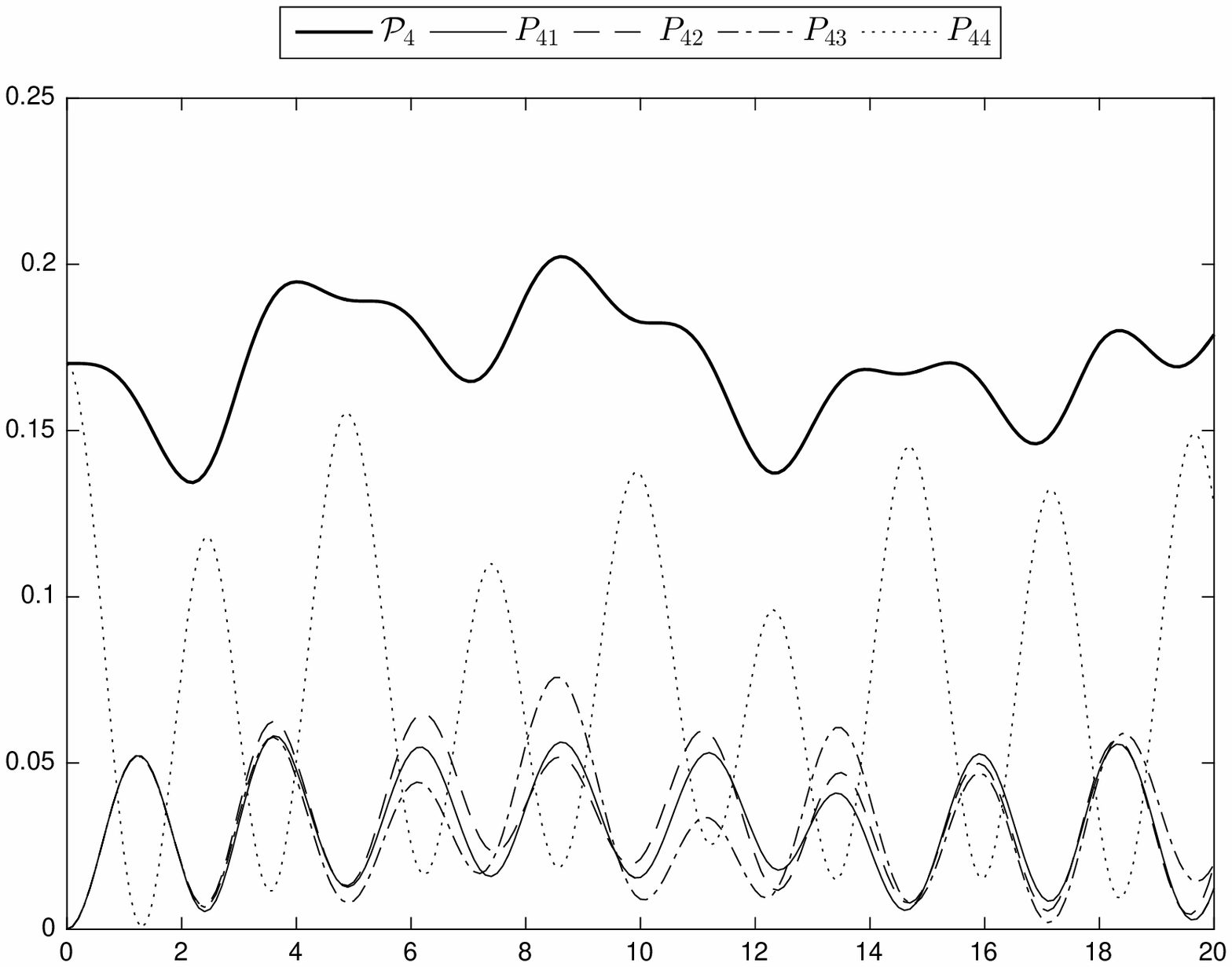}}
\caption{\footnotesize{\label{fig3} Dynamics of political parties affected by turncoat--like behaviors: the trend plotted in \ref{a2} captures the feature of the ``jump on the bandwagon''. The frames \ref{b2}, \ref{c2}, \ref{d2}, \ref{e2} show the fluctuations of the densities of the various subgroups of each party. The $x$--axis is scaled to the number of quarters in a five--year term of office; the rule $\rho^\prime$ is applied with $\tau=1$.
The plotted data are normalized to the sum of the initial densities.}}
\end{center}
\end{figure}

In this Section, we present an application of the model discussed above with $N=4$, whereupon $17$ 
compartments have to be considered.
The four parties $\mathcal{P}_i$ ($i=1,\ldots,4$) are intended to represent two more or less moderate groups ($\mathcal{P}_1$, $\mathcal{P}_2$) and two highly unfair ones ($\mathcal{P}_3$, $\mathcal{P}_4$). Different assignments of values to the set of parameters entering the Hamiltonian, apart from the initial datum, of course play a significant role in the evolution of the system giving rise to different dynamical behaviors. The key adopted here lies in the ideological and attitudinal interpretation of the political fringes; this means that the values of the inertia parameters (related to the tendency of each party to keep the same amount of members during the evolution) should be negatively correlated to those of the interaction parameters  (representing the disloyalty to one's own group and responsible for the internal/external flows), 
namely politicians belonging to parties with high inertia are less prone towards changes of party, whereas high interaction parameters 
are associated with low inertia parameters.  Notice that, in order to make the Hamiltonian operator self--adjoint, the parameters $\mu_{ij}$ ($i,j=1,\ldots,4$, $j>i$) definitely influence both the operator $P_{ij}$ and the operator $P_{ji}$.

Consider the $20$--quarter interval $[0, 20]$ corresponding with a $5$--year term of office, and split it into $n=20$ subintervals of equal length $\tau=1$.
The rule modifies at fixed times $k\tau$ ($k=1,\ldots,19$) the parameters $\omega_{ij}$ ($i,j=1,\ldots,4$), $\omega_5$, and $\lambda_{ij}$ ($j\neq i$) in \eqref{eq:Ham_turn} according to the variations of the densities of the groups in the time interval $[(k-1)\tau,k\tau]$. 
More precisely, at each step $k$, we compute the variations 
\begin{equation}
\begin{aligned}
\delta_{i}&=\mathfrak{p}_{i}(k\tau)-\mathfrak{p}_{i}((k-1)\tau),\quad i=1,\ldots,4,\\
\delta_5&=m(k\tau)-m((k-1)\tau).
\end{aligned}
\end{equation}

The first rule $\rho$ taken into account consists in the replacements
\begin{equation}
\label{eq:rho}
\left\{
\begin{aligned}
&\omega_{ii}\stackrel{\rho}{\longrightarrow}\omega_{ii} (1+\delta_{i}),\\ 
&\omega_{ij}\stackrel{\rho}{\longrightarrow}\omega_{ij} (1-\delta_{i}),\\
&\omega_5\stackrel{\rho}{\longrightarrow}\omega_5(1+\delta_5),\\
&\lambda_{ij}\stackrel{\rho}{\longrightarrow}\lambda_{ij} (1-\delta_{i}),
\end{aligned}
\right.
\qquad i,j=1,\ldots,4,\quad j\neq i.
\end{equation} 
The conditions \eqref{eq:rho} are intended to account for a reasonable reinforcement of the moderate parties by modifying, according to the evolution of the various groups, the tendency of the loyal fringes to remain constant in time in opposite way to the inertia of the disloyal subgroups and to the strength of the internal interactions (see Figure~\ref{fig2}). 

The rule $\rho$ may be improved in order to describe the feature related to the opportunistic attitudes of unfair politicians, called ``jump on the bandwagon'', consisting in supporting the movement that is seen to have become successful. For this aim, a second rule $\rho^\prime$ has been considered. The set of conditions $\rho^\prime$ acts as $\rho$ on all the parameters $\lambda_{ij}$ and on the parameters $\omega_{ij}$ with $i>1$, while the conditions modifying the inertia of the leading party now read
\begin{equation}
\label{eq:rhoprime}
\left\{
\begin{aligned}
&\omega_{11}\stackrel{\rho^\prime}{\longrightarrow}\omega_{11} (1+\Delta),\\ 
&\omega_{1j}\stackrel{\rho^\prime}{\longrightarrow}\omega_{1j} (1-\Delta),\qquad j=2,3,4,
\end{aligned}
\right.
\end{equation}
where $\displaystyle{\Delta=\max_{1\leq i \leq 4} \delta_i}$. Since the system under consideration is conservative, at any given time the variations $\delta_i$ will never be all negative, so that $\Delta>0$. This results in a suppression of the internal disloyalty and in a consolidation of the party $\mathcal{P}_1$ up the more its density grows (also due to the support arising from the  turncoats coming
from other groups), as shown in Figure~\ref{fig3}.

The numerical simulations in Figures~\ref{fig2} and \ref{fig3} have been produced by choosing the same initial values for the parameters, say 
\begin{equation}
\begin{aligned}
&\omega_{11}=0.7,\quad \omega_{12}=0.65,\quad \omega_{13}=0.65,\quad \omega_{14}=0.65,\\
&\omega_{21}=0.6,\quad \omega_{22}=0.65,\quad \omega_{23}=0.6,\quad \omega_{24}=0.6,\\
&\omega_{31}=0.4,\quad \omega_{32}=0.4,\quad \omega_{33}=0.45,\quad \omega_{34}=0.4,\\
&\omega_{41}=0.3,\quad \omega_{42}=0.3,\quad \omega_{43}=0.3,\quad \omega_{44}=0.35,\quad \omega_5=0.2,\\
&\lambda_{12}=\lambda_{13}=\lambda_{14}=0.35,\quad \lambda_{21}=\lambda_{23}=\lambda_{24}=0.4,\\ 
&\lambda_{31}=\lambda_{32}=\lambda_{34}=0.6,\quad \lambda_{41}=\lambda_{42}=\lambda_{43}=0.7,\\ 
&\mu_{12}=0.1,\quad \mu_{13}=\mu_{14}=0.15,\quad \mu_{23}=\mu_{24}=0.2,\quad \mu_{34}=0.25,\\ 
&\nu_{12}=\nu_{13}=\nu_{14}=\nu_{21}=\nu_{23}=\nu_{24}=0.1,\\ 
&\nu_{31}=\nu_{32}=\nu_{34}=0.15,\quad \nu_{41}=\nu_{42}=\nu_{43}=0.2,\\
\end{aligned}
\end{equation}
and by setting the initial densities for the compartments equal to
\begin{equation}
\begin{aligned}
&p_{11}(0)=0.45,\quad p_{22}(0)=0.10,\quad p_{33}(0)=0.40,\quad p_{44}(0)=0.05,\\ 
&p_{ij}(0)=m(0)=0,\quad i,j=1,\ldots,4,\quad j\neq i,
\end{aligned}
\end{equation}
and 
\begin{equation}
\begin{aligned}
&p_{11}(0)=0.80,\quad p_{22}(0)=0.70,\quad p_{33}(0)=0.45,\quad p_{44}(0)=0.40,\\
&p_{ij}(0)=m(0)=0,\quad i,j=1,\ldots,4,\quad j\neq i,
\end{aligned}
\end{equation}
respectively. In the last case, since the initial densities do not sum up to 1,  the densities of the various compartments plotted in Figure~\ref{fig3} are normalized.

A model belonging to the general one considered in Section~\ref{sect2} has been investigated in 
\cite{DSO_turncoat}, where, by suitably identifying the groups in both houses of the Italian Parliament 
of the XVII Legislature, a satisfactory fit with the real data of changes of side has been obtained. For that case study, the definition of the rule was somehow phenomenological, in the sense that it was strongly related to the dynamics observed inside the Italian political system during the period under study. In this paper, the disquisition is carried out in very general terms, and the two definitions of the rule seem us to be the most reasonable alternatives in this context. 

In conclusion, the operatorial approach joined with the introduction of specific rules, suitable to include in the dynamics meaningful  effects occurring during the evolution of the system, allows us to build in a straightforward way general deterministic mathematical models that, despite their simplicity, are able to produce interesting results in several contexts \cite{BDSGO_Hrho2016}.

\section*{Acknowledgments}
The research was partially funded by the Ph.D. School in Mathematics and Computer Science of the University of Catania.


\medskip

\begin{thebibliography}{99}

\bibitem{Johnson_Book} P.~E.~Johnson, Formal theories of politics. Mathematical modelling in political science. Elsevier Ltd, Pergamon (1989)

\bibitem{Lichtenegger} K.~Lichtenegger,  T.~Hadzibeganovic,  The interplay of self--reflection, social interaction and random events in the dynamics of opinion flow in two--party democracies, Int. J. Mod. Phys. C, \textbf{27} (2012)

\bibitem{Misra} A.~K.~Misra, A simple mathematical model for the spread of two political parties, Nonlin. Anal.: Model. Control, \textbf{17}, 343--354 (2012)

\bibitem{Sened} I.~Sened,  A model of coalition formation: theory and evidence, The Journal of Politics, \textbf{58}, 350--372 (1996)

\bibitem{Khrennikova_Haven} P.~Khrennikova,  E.~Haven,  A.~Khrennikov, An application of the theory of open quantum systems to model the dynamics of party governance in the US political system, Int. J. Theor. Phys., \textbf{53}, 1346--1360 (2014) 

\bibitem{Khrennikova} P.~Khrennikova, Quantum dynamical modeling of competition and 
cooperation between political parties: the coalition and non--coalition equilibrium model, J. 
Math. Psychol., \textbf{71}, 39--50 (2016)

\bibitem{Merzbacher} E.~Merzbacher, Quantum mechanics. J. Wiley \& Sons, New York (1970)

\bibitem{Roman} P.~Roman,  Advanced quantum mechanics. Addison--Wesley, New York (1965)

\bibitem{BagarelloBook} F.~Bagarello, Quantum dynamics for classical systems: with 
applications of the number operator. J. Wiley and Sons, New York (2012)

\bibitem{BagarelloOliveriMigration2013} F.~Bagarello, F.~Oliveri,   
An operator description of 
interactions between populations with applications to migration, Math. Mod. Meth. Appl. Sci., \textbf{23}, 471--492 (2013)

\bibitem{BagarelloHaven2014a} F.~Bagarello, E.~Haven, The role of information in a 
two--traders market, Physica A: Stat. Mech. Appl., \textbf{404}, 224--233 (2014)

\bibitem{BagarelloHaven2014b} F.~Bagarello, E.~Haven,  Toward a formalization of a two 
traders market with information exchange, Physica Scripta, \textbf{90} (2014)

\bibitem{BagarelloOliveriEco2014} F.~Bagarello, F.~Oliveri, Dynamics of closed ecosystems 
described by operators, Ecol. Model., \textbf{275}, 89--99 (2014)

\bibitem{BagarelloGarganoOliveriEscape2015} F.~Bagarello, F.~Gargano, F.~Oliveri,
A phenomenological operator description of dynamics of crowds: escape strategies, Appl. Math. Model., \textbf{39}, 2276--2294 (2015)

\bibitem{BagarelloCherubiniOliveriDesert2016} F.~Bagarello, A.~M.~Cherubini, F.~Oliveri, 
An operatorial description of desertification, SIAM J. Appl. Math., \textbf{76}, 479--499 (2016)

\bibitem{DiSalvoOliveriRM2016} R.~Di~Salvo, F.~Oliveri, An operatorial model for long--term survival of bacterial populations, Ricerche di Matematica, \textbf{65}, 435--447 (2016)

\bibitem{DiSalvoOliveriAAPP2016} R.~Di~Salvo, F.~Oliveri, On fermionic models of a closed ecosystem with application to bacterial populations, AAPP--Phys. Math.  Nat. Sci., \textbf{94}, No. 2, A5, doi: 10.1478/AAPP.942A5 (2016)

\bibitem{BagarelloAlliance2015} F.~Bagarello, An operator view on alliances in politics, SIAM J. Appl. Math., \textbf{75}, 564--584 (2015)

\bibitem{BagarelloHavenAlliance2016} {F.~Bagarello, E.~Haven},  {First results on applying a non--linear effect formalism to alliances between political parties and buy and sell dynamics}, Physica A, \textbf{444}, 403--414 (2016)

\bibitem{game1} M.~Peterson, The Prisoner's Dilemma. Cambridge University Press (2015)

\bibitem{game2} J.~Rothe, Economics and Computation: An Introduction to Algorithmic Game Theory, Computational Social Choice, and Fair Division. Springer-Verlag Berlin Heidelberg (2016)

\bibitem{DSO_turncoat} R.~Di Salvo, F.~Oliveri, An operatorial model for complex political system dynamics, Math. Meth. Appl. Sci., doi:10.1002/mma.4417 (2017)

\bibitem{DSGO_voters} R.~Di Salvo, M.~Gorgone, F.~Oliveri, $(H,\rho)$--induced 
political dynamics: facets of the disloyal attitudes into the public opinion, Int. J. Theor. Phys., \textbf{56}, 3912--3922 (2017)


\bibitem{BDSGO_QGOL2016} F.~Bagarello, R.~Di~Salvo, F.~Gargano, F.~Oliveri,  
$(H,\rho)$--induced dynamics and the quantum game of life, Appl. Math. Model., \textbf{43}, 15--32 (2016)

\bibitem{BDSGO_Hrho2016} F.~Bagarello, R.~Di~Salvo, F.~Gargano, F.~Oliveri, 
$(H, \rho)$--dynamics and large time behaviors, Submitted (2016)

\end{thebibliography}
\end{document}